\documentstyle[12pt]{article}

\begin{document}

\begin{center}
{\large\bf Black Hole Entropy in $1+1$ Dimensions from a Quasi-Chern
Simons term in a Gravitational Background}

\vspace{1cm}
{\sl Carlos Pinheiro}

\footnotesize{Departamento de F\'{\i}sica, CCE\\[-2mm]
Universidade Federal do Esp\'{\i}rito Santo -- UFES\\[-2mm]
Av. Fernando Ferrari S/N, Campus Goiabeira\\[-2mm]
29060-900 Vit\'oria, ES -- Brazil \\[-2mm]
fcpnunes@@cce.ufes.br/maria@@gbl.com.br}

and 

{\sl F.C. Khanna}\\
\footnotesize{Theoretical Physics Institute, Dept. of Physics\\
University of Alberta,\\
Edmonton, AB T6G2J1, Canada\\
and\\
TRIUMF, 4004, Wesbrook Mall,\\
V6T2A3, Vancouver, BC, Canada.
khanna@@phys.ualberta.ca}
\end{center}

\normalsize

\vspace{0.5cm}

\begin{abstract}
We introduce a 'quasi-topological` term [1] in $D=1+1$
dimensions and the entropy for black holes is calculated [2]. The source of
entropy in this case is justified by a non-null stress-energy tensor.

\end{abstract}

\newpage
\setcounter{page}{1}
\section{Introduction}
\paragraph*{}

It is well known that higher derivative Chern-Simons extensions in a
background of gravitation [1] in $D=1+2$ dimensions are possible.
Such action has some pecularities compared to the usual
Chern-Simons gravitational or Abelian Chern-Simons theory. Perhaps
the most interesting fact associated with an extension for
Chern-Simons theory in $D=1+2$ is that the energy-momentum tensor is
non zero. Considering the action given by Deser-Jackiw [1] it is not
difficult to calculate the entropy for black holes in $D=1+2$
dimensions [6].

Despite the fact that the Chern Simons theory appears only  in odd
dimensions [5,3] we attempt to write here an analogous action in a
background of gravitation in $D=1+1$ dimensions and compute the
entropy of black holes in such a case.

This action in $D=1+1$ dimension may be called a ``quasi-topological
action as Chern-Simons'' in the same sense as suggested earlier
[1,3,6,8]. 

The quantities such as Hawking's temperature, inverse temperature and
entropy correction are shown as in the corresponding case given
before [2,6].

The ``quasi topological action'' is different from the one suggested
earlier [2] and depends locally on the potential vector in flat space
time. 

Let us start by writing the functional integral following the analogy
of the Euclidean field theory and statistical mechanics as
\begin{equation}
Z_T = \int \ {\cal D}g \ e^{\displaystyle{-(I[g,\varphi ] + I[f,g])}}
\end{equation}
where 
\begin{equation}
I[g,\varphi ]=\frac{1}{4G} \int \ d^2x \
\sqrt{-g} \ e^{-2\varphi} \left[R+4(\nabla\varphi )^2 + 4\lambda^2\right]\ . 
\end{equation}
Here $I[g,\varphi ]$ is the two dimensional  dilaton action motivated by the
string theory [7] and
\begin{equation}
I[f,g] = \int d^2x \ \varepsilon^{\mu\nu} f_{\mu}\Box^2 f_{\nu}
\end{equation} 
It's a ``quasi topological term'' in $1+1$ dimensions. In equation
(3) $f_{\mu}$, is a covariant vectors with
$f_{\mu}$ given as
\begin{equation}
f_{\mu} = g^{-1/2} g_{\mu\beta} \varepsilon^{\beta\lambda}
A_{\lambda} \ .
\end{equation}
The quantities $R$, $\varphi$, $g,\varepsilon^{\beta\lambda}$ and
$A_{\lambda}(x)$ are respectively scalar curvature, dilaton field,
determinant of the metric, the Levi-Cevita tensor
$\varepsilon^{01}=+1$ and the vector potential respectively.

A general solution  of the Einstein's equation is parametrized by one
constant [7] and it is given by
\begin{equation}  
ds^2 = - A \ dt^2 + \frac{dr^2}{A}
\end{equation}
where
\begin{equation}
A = 1-\frac{M}{\lambda} \ e^{-\alpha \lambda r}
\end{equation}
and
\begin{equation}
\varphi = -\lambda r \ .
\end{equation}

The metric describes a static, asymptotically flat two dimensional
black hole with event horizon at $r=r_+$ where
\begin{equation}
r_+ = \frac{1}{2\lambda} \ln \ \left(\frac{M}{\lambda}\right) \ .
\end{equation}
with the parameter $M$ being identified with the mass of a black hole.

Following [2] we can find the Hawking's temperature $\alpha$, inverse
temperature as well the Euclidean time period $\beta$, and entropy $S$
as
\begin{eqnarray}
&& \alpha = \frac{df(r)}{dr}\Big|_{r=r_+}\ , \\
&& \nonumber \\
&& \beta = \frac{2\pi}{\alpha} \ ,\\
&& \nonumber \\
&& S = \frac{\tilde{A}_0}{4} 
\end{eqnarray}
with $f(r)$ being equal to $A$ and $\tilde{A}_0$ meaning the area of
the event horizon in $D=1+1$ dimensions and $\tilde{A}_0 = r_+$
respectively. These quantities are then given as
\begin{eqnarray}
&& \alpha = 2\lambda \ , \nonumber \\
&& \beta = \frac{\pi}{\lambda }\ , \\
&& S = \frac{1}{8\lambda} \ \ln \left(\frac{M}{\lambda}\right)\ . \nonumber
\end{eqnarray}
It is usual to redefine the Hawking's temperature as
\begin{equation}
T_H = \frac{1}{\beta} \ ,
\end{equation}
thus, in this case the temperature is given as
\begin{equation}
T = \frac{\lambda }{\pi} \ .
\end{equation}
The equation (3) may be expressed as 
\begin{equation}
I[f,g] = \int d^2x (f_0\Box^2 f_1-f_1\Box^2 f_0) 
\end{equation}
using equation (4), the two contributions $f_0$ and $f_1$ are given as
\begin{eqnarray}
&& f_0 = \left(1-\frac{M}{\lambda} \ e^{-2\lambda r}\right) \ A_1 (x)
\nonumber \\
&& \\
&& f_1 = - \left(1-\frac{M}{\lambda} \ e^{-2\lambda r}\right)^{-1} 
\ A_0 (x)  \ . \nonumber
\end{eqnarray}
It is important to remember that equation (15) was obtained
considering only the static configuration for $f_0$ and $f_1$ fields.

Then we have
\begin{equation}
Z_T \cong e^{\displaystyle{f_0\Box^2f_1-f_1\Box^2f_0}} \ \int {\cal D} g \
e^{\displaystyle{-(I[g,\varphi ] + I[f,g])}} 
\end{equation}
where only two terms in integrand (15) are given as
\begin{equation}
f_0\Box^2 f_1 - f_1\Box^2 f_0 \cong \Box \frac{4M \ 
e^{-2\lambda r}}{\left(1-\frac{M}{\lambda} \ e^{-2\lambda r}\right)} 
\ F(A_1,A_0) 
\end{equation}
and
\begin{equation}
\ \ F(A_1,A_0) = A_1 \left(\frac{\partial A_0}{\partial
r}\right) + A_0 \left(\frac{\partial A_1}{\partial
r}\right) \ .
\end{equation}
Considering now the following approximation:
$f_0\Box^2 f_1 - f_1\Box^2 f_0 \simeq \Box \left(f_0\Box f_1-f_1\Box f_0\right)$
where we wish to imply that the operator $\Box$ will act only on the
second part of each term of the expression  
$f_0\Box f_1-f_1\Box f_0$, i.e, it acts on $\Box f_1$ and $\Box f_0$ only.

We approximate the partition function associated with the ``quasi
topological term'' as
\begin{equation}
Z^{Chern}_{Simons} \simeq  \ e^{\displaystyle{f_0\Box^2f_1-f_1\Box^2f_0}}
\end{equation}
\begin{equation}
Z^{Chern}_{Simons} = \ e^{\displaystyle{\Box\left(\frac{4\pi F(A_1,A_0)}{\beta e^{2\lambda
r}-\pi /\lambda}\right)}}
\end{equation}

The contribution to entropy is found from 
\begin{equation}
S  = \ln \ Z- \beta \ \frac{\partial}{\partial\beta} \ \ln Z
\end{equation}
and we get
\begin{equation}
S \approx \Box \left[\frac{4\pi \ F(A_0,A_1)}{\left(\beta \ e^{2\lambda
r}-\frac{\pi}{\lambda}\right)}  + \frac{4\pi\beta \ F(A_0,A_1)
e^{2\lambda r}}
{\left(\beta \ e^{2\lambda
r}-\frac{\pi}{\lambda}\right)^2}\right] 
\end{equation}
The average energy is obtained from
\begin{equation}
M = -\frac{\partial}{\partial\beta} \ln \ Z
\end{equation}
and is 
\begin{equation}
M \cong \Box\left[\frac{4\pi \ F(A_1,A_0) e^{2\lambda r}}{\left(\beta \ e^{2\lambda
r}-\frac{\pi}{\lambda}\right)}\right]
\end{equation}

It is assumed that all fields at infinity go to zero i.e.
$F(A_0,A_1)\rightarrow 0$ when $r\rightarrow \infty$.

The entropy, (23), and mass, (25), are zero in that case and we can
recover the results (12) for Einstein-Hilbert theory only. For
$r=r_+$ the equations (23) and (25) diverge then we have a  
mechanism for generation of mass and entropy. Finally, for $0<r<r_+$ a
positive value for $S$ is obtained. 

In the general case, however, the total entropy is 
\begin{equation}
S_T \sim  S^{Einstein}_{Hilbert} \oplus S^{Chern}_{Simons}
\end{equation}
with
$S^{Einstein}_{Hilbert}$ given by (12) and
$S^{Chern}_{Simons}$ given by eq. (23). 

Again the source of entropy from action (3) is traced to the fact that the
energy momentum tensor is not zero but is given by
\begin{equation}
T^{\mu\lambda} = \frac{2\Box}{g^2}\ \varepsilon^{\nu k}
\varepsilon^{\theta\gamma} \varepsilon^{\xi\delta}
g^{\mu\lambda} g_{k\theta}
g_{\nu\xi} A_{\gamma}A_{\delta}
\end{equation} 
or using (4) it can be written as 
\[
T^{\mu \lambda}=\frac{2\Box}{g}\ g^{\mu \lambda}f_kf^k
\]
Then $T^{00}\sim g^{00}$ and at $r=r_+$, the stress energy momentum
tensor diverges. On the other hand, in the limit $r\rightarrow
\infty$, $T^{\mu\lambda}$ is null due to our assumption that all
fields are zero at infinity; so the metric (5) reduces to the flat
space time with eq. (4) going to zero along with eq. (3).

The conservation of $T^{\mu\lambda}$ on-shell is easily checked with the
equation of motion given as
\begin{equation}
\varepsilon^{k\nu}\varepsilon^{\theta\gamma}\varepsilon^{\xi \delta}
g_{k\theta}g_{\nu \varepsilon}A_{\gamma}A_{\delta} = 0
\end{equation}
or still yet
\begin{equation}
\varepsilon^{\nu}_{\theta}\varepsilon^{\theta\gamma}\varepsilon^{\delta}_{\nu}
A_{\gamma}A_{\delta} = 0
\end{equation}
and finally using again (4) it is given as 
\[
\varepsilon^{k\nu}f_kf_{\nu}=0
\]
and thus, independent of the metric. The complete metric dependence
is on the field $f_{\mu}$ given by eq, (4).

Hence, the action (3) or (15) is clearly not completly a topological
action as in [5,4] but is a ``quasi-topological in the sense that the
metric dependence is entirely contained in $f_{\mu}$ in the same
way as [1,6,8]. It is only with this meaning that we are calling the
action (3) ``A quasi-chern-simons action'' or ``quasi-topological
term''. The action (3) has the same form as that of Jackiw's action in [1],
but there taking the limit of $g_{\mu\nu}$ for $\eta_{\mu\nu}$, flat
space time, the higher derivative Chern-Simons extension can be
written as $I_{ECS}\sim \int d^3x \ f_{\alpha} \partial_{\beta}
f_{\gamma}\varepsilon^{\alpha\beta\gamma}$ and $f^{\alpha} =
\displaystyle{\frac{1}{2}} \varepsilon^{\alpha\mu\nu}F_{\mu\nu}$
where the extension depends locally on the field strength and not on
the  vector potential. In our present case, the equation  (4) is
reduced to $f^{\alpha} = \varepsilon^{\lambda\beta}A_{\beta}$ and
then the action (3) depends
locally on the  vector potential directly, while the 
equation of  motion is independent of metric, thus, our case being
different than that of [1]. However, the best analogy with $D=2+1$ we
can do is through considering the equation (3) and some
quasi-topological aspects such in [1,3,8].

Hence, for $D=1+1$ dimension   we can't write an exact topological
Chern-Simons action but one can evaluate some contribution for
entropy of black holes by constructing an action analogously to [1]
and improving the same approach as in [2].

\section*{Conclusions and Comments}
\paragraph*{}

We had introduced a ``quasi topological action'' as in [1] and
with an appropriate definition of a vector in $D=1+1$ dimension. The
contribution to entropy of black holes in two dimensions is found.

The correction for average energy has been written as a function of
$F(A_0,A_1)$ with the assumption that all fields at infinite are zero.

The contribution to entropy is attributed to a non-null energy momentum
tensor (27). We justify the source of entropy  as the stress-energy tensor.

\section*{Acknowledgements}

\paragraph*{}
I would like to thank the Department of Physics, University of
Alberta for their hospitality. This work was supported by CNPq
(Governamental Brazilian Agencie for Research.

I would like to thank also Dr. Don N. Page for his kindness and attention
with  me at Univertsity of Alberta.

%\newpage

\end{document}